\documentclass{article}
\pdfoutput=1
\usepackage[utf8]{inputenc}

\title{NSWTransport}
\author{ }
\date{February 2017}

\usepackage{graphicx}
\usepackage{xspace}
\usepackage[in]{fullpage}
\usepackage{amsmath}
\usepackage{amsfonts}
\usepackage{amssymb}
\usepackage{amsthm}
\usepackage{xargs}   
\usepackage{color}
\usepackage[pdftex,dvipsnames]{xcolor} 

\newcommand{\alg}{Second Algorithm\xspace}
\newcommand{\term}[1]{\textbf{#1}\xspace}

\newcommand{\reals}{\ensuremath{\mathbb{R}}\xspace}
\newcommand{\naturals}{\ensuremath{\mathbb{N}}\xspace}
\renewcommand{\Pr}[1]{\ensuremath{\mathrm{Pr}\left({#1}\right)}\xspace}
\renewcommand{\v}[1]{\ensuremath{\mathbf{#1}}\xspace}

\theoremstyle{plain}
  \newtheorem{thm}{\protect\theoremname}
\theoremstyle{definition}
  \newtheorem{defn}[thm]{\protect\definitionname}
  \newtheorem{lem}[thm]{\protect\lemmaname}

\theoremstyle{plain}

\usepackage[english]{babel}
  \providecommand{\definitionname}{Definition}
  \providecommand{\lemmaname}{Lemma}
  \providecommand{\propositionname}{Proposition}
  \providecommand{\corollaryname}{Corollary}
  \providecommand{\remarkname}{Remark}
  \providecommand{\examplename}{Example}
\providecommand{\theoremname}{Theorem}

\newcommand{\eg}{\emph{e.g.},\xspace}
\title{Privacy Assessment of De-identified Opal Data: \\
A report for \emph{Transport for NSW}}
\date{\today}
\author{Chris Culnane, Benjamin I.~P. Rubinstein, Vanessa Teague \\
University of Melbourne \\
{\tt \{vjteague,  benjamin.rubinstein, christopher.culnane\}@unimelb.edu.au } }

\begin{document}

\maketitle

We consider the privacy implications of public release of a de-identified dataset of Opal card transactions. 
The data was recently published at {\tt https://opendata.transport.nsw.gov.au/dataset/opal-tap-on-and-tap-off}.
It consists of tap-on and tap-off counts for NSW's four modes of public transport, collected over two separate week-long periods.  The data has been further treated to improve privacy  by removing small counts, aggregating some stops and routes, and perturbing the counts.   This is a summary of our findings.

\subsection*{About the De-identified Data}
\begin{itemize}
    \item This version of the dataset can reasonably be released publicly.
    \item Many of the paired tap-on/tap-off events have been decoupled in the proposed release datasets. Removing the links between related tap-ons and tap-offs, such as those for the same trip, journey, or passenger, is critical to successful privacy protection. We recommend further decoupling in future, {\it e.g.,} Manly events. 
    \item The other significant privacy protection comes from the aggregation or suppression of small counts and sparsely patronised routes.
    \item It is still possible to detect the presence of a suspected individual or small group, with a small probability, in some unusual circumstances.  Note that this detects that \emph{someone} was present, without providing information about who it was.  These are probably not a matter of serious concern because of the very small probabilities involved---we estimate less than one in a thousand for small groups, and even less for individuals (about 2 in a million).  However, it is important to understand that this is possible and to make a risk assessment based on the likelihood in realistic cases.   We provide some specific examples and probability estimates in the main report.
    \item We would not recommend reusing this approach on more sensitive data releases such as trip or trajectory microdata. 
    \item It may be reasonable to release more weeks of data, with a careful assessment of how the risks increase as more data is available. 
\end{itemize}

\subsection*{About the De-identification Techniques}
Differential privacy is the gold standard for rigorous privacy protection---we support efforts to make open data provably differentially private.  However, it is important to understand that differential privacy does not imply perfect privacy protection, but rather the opportunity to quantify privacy loss.  We suggest the following improvements to the specific treatment of DP in this dataset:
\begin{itemize}
  \item The DP techniques and parameters should be made public, including the parameters for perturbing the totals.  
  
    This is good for utility because it allows those analysing the data to understand with what confidence their conclusions hold.  It is also good for privacy because it allows a rigorous risk assessment based on the degree of privacy protection, which is never perfect. 
    
  \item The perturbation parameters can be inferred from the data itself anyway, furthering the argument that they might as well be public.  Our estimates are in the full report.
  \item The DP treatment achieves at best $(\epsilon, \delta)$-differential privacy for a small, but non-zero constant $\delta$. This represents a weaker (less private) form of privacy than the $\epsilon$-differential privacy variant. The main reason for this is the decision not to perturb zero counts---we recommend that perturbation be extended to zeros in future.
  \item One dataset includes counts with both times and locations, another has only temporal data, and a third has only locations.  These three seem to have been derived independently from the same raw data. 
  Presenting three differently-treated versions of the same data may have unexpected implications for privacy. It would be better to derive the temporal data and the spatial data from a differentially-private version of the combined time and location data. 
  
\end{itemize}

The main report describes the specific inferences that can be made.
First we explain how it was possible to recover the parameters of the perturbations used for differential privacy.  This is not a problem, just a further reason to make those parameters public anyway.  The differential privacy framework does not rely on secrecy of these parameters.

We then use those parameters to quantify the likelihood of an attacker detecting the presence of individuals or small groups in the published data.  The probabilities are small, but these sorts of risks should be considered and recomputed before many more weeks of data are released.

Third, we discuss how some suppressed values could 
be recovered with reasonable accuracy based on calculating differences.  This demonstrates that independently treating the data in three different ways might inadvertently expose information. 

\newpage

\section{Overview of the Data and Why it Protects Privacy}
The main risk from transport data is that an attacker could use partial information about someone's travel patterns to re-identify their record and learn other information about their trip or journey.  For example, seeing where someone gets onto a train might, when linked with the raw data, expose where they get off.  If the raw data links events for the same trip, journey or Opal card, then that re-identification might expose other trips or journeys by the same person. 

The released dataset lists tap-on and tap-off counts from two non-contiguous weeks of data from the Opal public transport ticketing system.    
Effectively all trip information has been removed---there is (almost) no way to link different ends of the same trip, or different trips by the same person.  This means that partial information about a person's travel cannot be linked with the Opal data  to extract more information such as the other end of their trip or the locations of their other journeys.  The removal of these links is critical to good privacy protection, though of course it has a corresponding effect on utility because it prevents analysis of trips and journeys.

Another risk is that an attacker could use public data to detect the presence or absence of a suspected traveller at a particular place and time.  Although this risk is mostly mitigated in the dataset, it is not entirely eliminated. This report gives some specific examples with estimates of the (small) probability of detection, then some suggestions for improving the treatment to reduce the risk in future releases.

The dataset includes records for the four different modes of NSW public transport (train, bus, ferry and light rail), with times binned into 15 minute intervals. 
There are three different datasets with slightly different aggregation techniques---one contains time and location data, another only times and a third only locations.  

Each count is perturbed by a random value chosen from the Laplace distribution paramaterised by a privacy parameter $p$.  We take this to mean that the count in the dataset is 
$$ c = c_{\text{raw}} + L$$
for some true value $c_{\text{raw}}$ and noise random variable $L\sim \textit{Lap}(0, p)$.  The probability of perturbation level $x$ is 
$$ \Pr{ L = x} =  \frac{1}{2p}\exp(-|x|/p)\enspace. \ $$

Counts less than 18 after perturbation have been removed.  In some datasets, an even higher threshold has been applied.

Further details about the data treatment are in a report prepared by Data61.

\section{Recovering Perturbation Parameters by Analysis of Differences} \label{sec:Recovering}

We can't immediately tell how much a number has been perturbed by.  However, we can get an estimate because in some instances we observe two different perturbed values which both started from the same raw number.

Notably, there are two ferry services which operate point-to-point, {\it i.e.}  they have a single start and end point.  Those services are the Manly Ferry, between Circular Quay Wharf 3 and Manly Wharf, and the Newcastle Ferry between Stockton Wharf and Queens Wharf. 

The Manly ferry has a duration of 30 minutes and is extremely popular. This results in many trips being distinguishable within the dataset, in which a tap on and a tap off can be paired. An added benefit of the Manly ferry route is that it is the only route that operates an automatic tap-off function, to improve speed of disembarking. This is important from an analysis perspective because it means everyone who tapped on will definitely tap off, whereas for other routes there may be a small number who forget to tap off. Using this property it is possible to look at paired tap on and tap offs. Two examples are given in Table~\ref{ferryExtract}.

\begin{table}[htbp]
\centering
\begin{tabular}{|l|r|l|l|l|r|}
\hline
ferry & 20160725 & on & 06:45 & Manly Wharf & 88 \\ \hline
ferry & 20160725 & off & 07:15 & Circular Quay No. 3 Wharf & 86 \\ \hline
\multicolumn{6}{c}{} \\ \hline
ferry & 20160725 & on & 07:00 & Circular Quay No. 3 Wharf & 33 \\ \hline
ferry & 20160725 & off & 07:30 & Manly Wharf & 32 \\ \hline
\end{tabular}
\caption{Manly Ferry Extract showing different perturbations of the same raw value.}
\label{ferryExtract}
\end{table}

\begin{figure}[htbp]
\centering
\includegraphics[width=\textwidth]{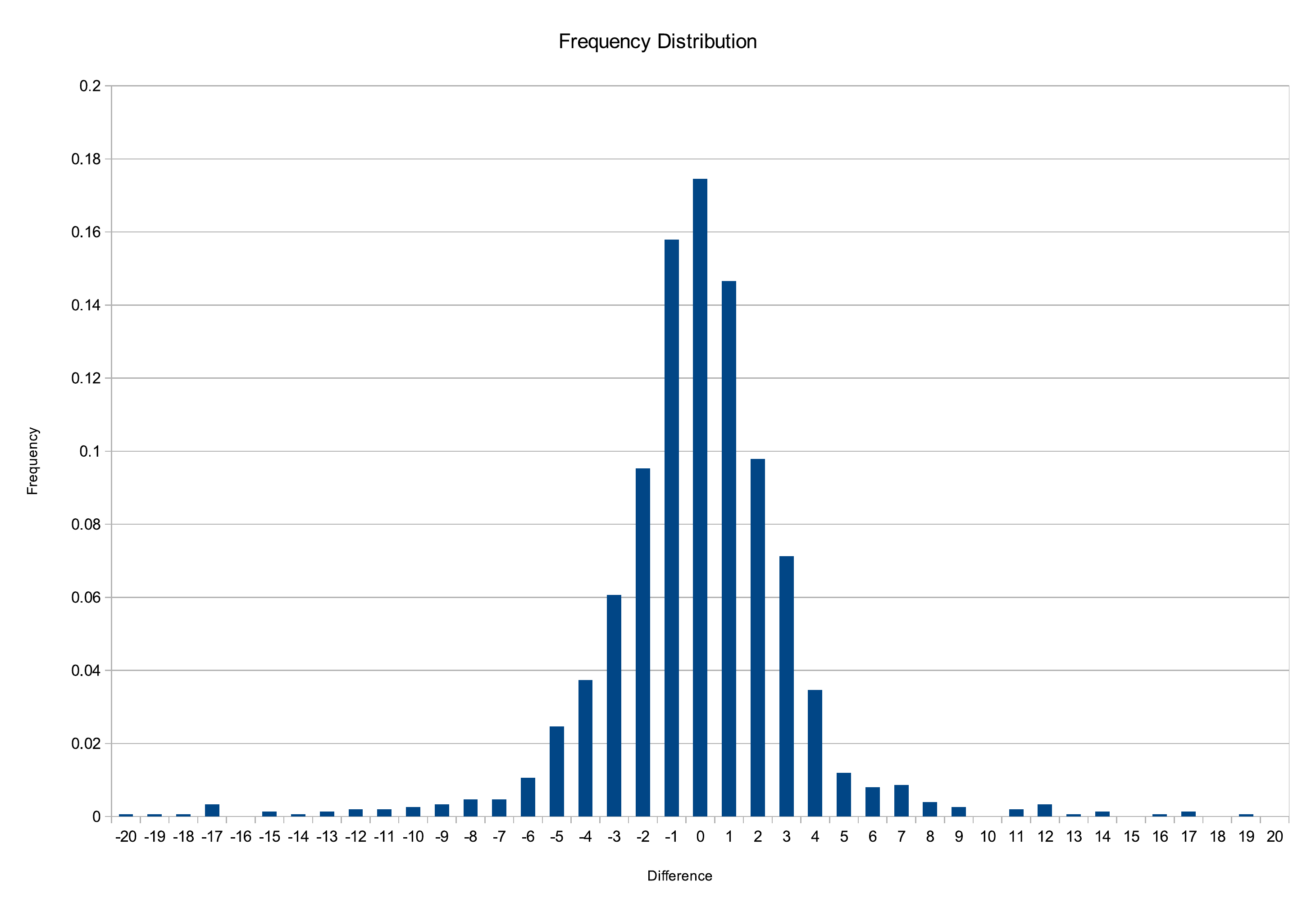}
\caption{Frequency Distribution of Manly Ferry Count Differences}
\label{fig:freqdist}
\end{figure}

In the raw data the number of tap ons and the equivalent tap offs must be exactly the same. In the released dataset we see small differences between these numbers because an independent randomly-chosen perturbation has been applied to each.
If we plot the frequency distribution of those differences we get the plot in Figure \ref{fig:freqdist}.  This is exactly the distribution we would expect from the difference between two Laplace distributions---the mathematical details are in Appendix~\ref{app:LaplaceDifferences}.  This is a strong indicator that the differences we are seeing are a result of the differentially-private algorithm. This allows us to estimate the perturbation parameter $p$ as approximately 1.4.

Whilst we don't learn much directly from recovering the parameter, it does shed some light onto other areas of the dataset and raises a number of questions around the differential-privacy assumptions. The most important of these is the assumption of independence of tuples within the dataset. When this assumption fails it can result in the privacy guarantees being lower than expected \cite{kifer2011no}, \cite{liu2016dependence}. In other words, the privacy budget could be exhausted more quickly, or possibly even exceeded.  In this example, the two different records of the number of ferry passengers allow a more accurate guess of the raw value than one record alone would give.

Recovering this parameter should not have an impact on privacy, and it would be perfectly reasonable to publish these sorts of algorithmic details.  Indeed this demonstrates that nothing is gained by hiding them, because they can often be recovered.  Publishing the parameters has a real benefit to utility, too, because it allows data analysts to assess the confidence of their conclusions.

\section{Recommendations to Achieve Strong Differential Privacy}  \label{sec:DPImpossible}
In this section we should how an attacker could, with very small probability, detect the presence of a suspected individual or small group. These are probably not a matter of serious practical concern because of the very small probabilities involved---about 2 in a million for individuals.  However, it does mean that the dataset does not technically meet the precise definition of strong Differential Privacy.  We explain why, and show how this can be easily corrected for future releases.

The released dataset consists of a number of records of certain types (such as tap ons at a particular time and location). Data61's report addendum describes the ``\alg'' (also labelled as Algorithm 1 in Section 2, Page 4) for perturbing the counts of each type of record. The addendum asserts that the algorithm preserves differential privacy, according to convention implying the strong form of $\epsilon$-differential privacy.  However, strictly speaking it doesn't meet this definition.  We explain why it doesn't and how it can be easily corrected.

The algorithm follows standard techniques but has a subtle error: when a count is zero, it is not perturbed.  This means that if an adversary observes a non-zero perturbed value, it has certainly been derived from a non-zero raw value.  
Put plainly, an output above the algorithm's suppression
threshold can be used to rule out some datasets---those with a zero count at that point. 

This is easy to correct---simply include zero counts in the perturbations applied to all the other counts.  In other words, remove from Algorithm~1 the special case in line 3, ``if c is zero, continue.''  

The effect of the error is greatly reduced by the suppression of small values---a small nonzero value would have to be perturbed up to 18 in order to be detected.  This is the main reason that the error doesn't represent a serious issue for privacy---a large enough perturbation for detection would be very improbable.  However, if the same algorithm is used for datasets that do not suppress small values, the correction should be made.

A concrete example would be a very infrequently used stop, for example in an outer urban area at 5am.  Suppose the attacker knows that there is only one person who ever uses that stop at that time, and wants to know whether she used it on a particular date.  If the dataset contains a zero count, the attacker doesn't learn anything: it might have been a true zero, or it might have been a one perturbed down to zero.  However, if the attacker observes an 18, it could not have originated as a zero.  It must have been (at least) a one in the raw data, so the person must have been there.  This is rather a far-fetched example, and the probabilities are extremely small, so in practice this is not a serious problem for this data.

However, being able to detect with certainty the presence of at least one person (even with a small probability) implies that the algorithm does not achieve the strong form of differential privacy.
We prove this formally in Appendix~\ref{app:not-edp}. Moreover, while the \alg may achieve the weaker $(\epsilon,\delta)$-differential privacy, we prove a lower bound on the $\delta$ possibly achievable in terms of whatever perturbation parameter $p$ the \alg is run with.
Usually $(\epsilon,\delta)$-differential privacy requires $\delta$ to be negligible in the size of the database.  That is not the case here---$\delta$ is a (very small) constant value.

\begin{thm}\label{thm:not-edp}
Irrespective of chosen threshold, the \alg does not preserve $\epsilon$-differential privacy for any $\epsilon$.
It cannot preserve $(\epsilon,\delta)$-differential privacy for $\delta<\exp(-17/p)/(2p)$.
\end{thm}

We can now use the estimate from Section~\ref{sec:Recovering} that $p \approx 1.4$. This then implies that $\delta$ must be at least $\exp(-17/1.4)/(2.8) \approx 0.000002$.  This is the probability of detecting, for certain, an individual at a stop that otherwise had a count of zero.  The probabilities for small groups are slightly larger---the same estimation gives a probability of about 0.00004 for detecting a group of 5 and 0.005 for a group of 12.\footnote{These numbers depend on our estimate of $p$.  If a precise version of $p$ is known then it can be used to calculate more accurate bounds.}

Security and privacy is full of examples of  subtle errors with serious consequences.  This issue was hard to detect, but is easy to fix.  This shows the value of careful review---making the algorithm and its analysis public would increase the likelihood that other errors and weaknesses could be found and corrected.

\section{Estimating Suppressed Values from Multiple Queries}
The raw data is presented in three different datasets for each mode of transport: one dataset includes both times and locations, a second aggregates all the times for a given location, and a third aggregates locations and lists only times.  When a stop with a small count is suppressed from the time-and-location dataset, it is sometimes possible to estimate the missing value because it forms part of a total in the time-only dataset and has not been suppressed there.

This is equivalent to having run multiple queries on the same dataset. When we use the term ``query'' we regard aggregating mode, tap offs and time as being a distinct query from aggregating mode, location, tap offs and time. 
When the level of privatizing perturbation is not high (or suppression mitigates the level of differential privacy), then it can be possible to recover some suppressed information using the differences between the datasets that report the same information.  Although we don't recover the exact value, we recover a range of possible values within a confidence interval. 

In future it would be better to generate a differentially-private time-and-location dataset, perform suppression, and then derive the time-only and location-only datasets from it, not directly from the raw data.

\subsection{Spotting Secret Ferries Late At Night}

\begin{table}[htbp]
\centering
\begin{tabular}{|l|l|r|l|l|l|r|}
\hline
\textbf{Dataset} & \textbf{Mode} & \multicolumn{1}{l|}{\textbf{Date}} & \textbf{Type} & \textbf{Time} & \textbf{Location} & \multicolumn{1}{l|}{\textbf{Count}} \\ \hline
time-loc & ferry & 20160812 & off & 00:00 & Manly Wharf & 91 \\ \hline
time-loc & ferry & 20160812 & off & 00:00 & Circular Quay No. 3 Wharf & 41 \\ \hline
\multicolumn{7}{c}{} \\ \hline
time & ferry & 20160812 & off & 00:00 & - & 150 \\ \hline
\end{tabular}
\caption{Comparing different totals for the same raw data to estimate the suppressed value.}
\label{timelocaext}
\end{table}

Table \ref{timelocaext} contains an extract from two different datasets that report the same trips. The first two rows are from the time-and-location dataset, whilst the third row is from the time dataset. In the case of the former, they are the only records related to Ferry journeys made at that time on that date. As we can see the total number of people getting off at that time in the time-and-location dataset is 132. That value has been perturbed, so the raw value could be slightly larger or smaller than that. The total number of tap offs across the network at that time was 150 according to the time dataset. This is a sufficiently large  difference to indicate that the time dataset contains (nearly 18) additional records that have been suppressed from the time-and-location dataset. 

By referencing the timetables for the various ferry routes we can see that there are only a limited number of ferries that are timetabled to arrive in the 00:00 to 00:14 time window. Table~\ref{timetable} shows an extract from the ferry timetables. Only the F1 and F6 ferries are running at that time. We have already accounted for the F1 ferry in the time-and-location dataset. We can therefore by fairly certain that the remaining value belongs to the F6 ferry---unless the perturbations were so large as to account for the observed  difference. This leaves two possibilities, either Cremorne Point or South Mosman, the most likely is Cremorne Point, since for those tap off's to be in South Mosman it would require the ferry to arrive early, and for all those disembarking to move up the ramp to the Opal reader and tap off in an extremely short period of time. Therefore, with a reasonable degree of certainty, we can assume that those passengers in fact got off at Cremorne Point. 

Although this does not give us a precise count of Cremorne Point tap offs, it does allow us to estimate it.  A large value (nearly 18) is much more likely to have produced this observation than a small value (less than 5).  Details of the estimates are in Appendix~\ref{sec:confidenceEst}.

\begin{table}[htbp]
\centering
\begin{tabular}{|l|l|l|r|r|}
\hline
\textbf{Route} & \textbf{Origin} & \textbf{Destination} & \multicolumn{1}{l|}{\textbf{Departs}} & \multicolumn{1}{l|}{\textbf{Arrives}} \\ \hline
F1 & Manly & Circular Quay & 23:40 & 00:10 \\ \hline
F1 & Circular Quay & Manly & 23:45 & 00:15 \\ \hline
F6 & Circular Quay & Cremorne Point & 00:00 & 00:10 \\ \hline
F6 & Cremorne Point & South Mosman & 00:10 & 00:15 \\ \hline
\end{tabular}
\caption{Extract of Ferry Timetable}
\label{timetable}
\end{table}

\subsection{Timetable Extremes}
A very similar issue arises for sparsely patronised bus routes, which are also suppressed differently in different versions of the released data.
The process of binning, or combining, multiple points is a valid approach for protecting privacy. In the released dataset bus stops have been binned to their postcode. This is very effective during peak hours, when there are many buses and many bus stops active in any one postcode during any 15 minute time interval. However, when operating at the extremes of the timetable the frequency of buses and active bus stops dramatically reduces, leading to situations whereby the exact location of the bus stop and even the bus the passengers go onto is identifiable. This only has limited impact on privacy since low counts are still removed, but it does reveal some additional information. For an example of this we can look at the N70 night bus route on 2016/07/26. It leaves Penrith Interchange at 04:16. This is in the postcode of 2750. We don't see any time and location entries corresponding to that, so must assume that either no one got on the bus there or only a number below the threshold got on. Table \ref{tab:bus_extract} shows two rows for postcodes that count the Mount Druitt Station and Blacktown Station bus stops respectively. By comparing with the bus timetables covering the area we can be certain that these two rows refer the N70 that left Penrith at 04:16 and is headed to City Town Hall. In this instance we do not learn much more due to the time overlapping with the start of a the regular services shortly after 05:00 which acts to hide the trip. 

\begin{table}[htbp]
\begin{center}
\begin{tabular}{|l|r|l|l|r|r|}
\hline
bus & 20160726 & on & 04:30 & 2770 & 21 \\ \hline
bus & 20160726 & on & 04:45 & 2148 & 31 \\ \hline
\end{tabular}
\end{center}
\caption{Extract of time and location for night bus}
\label{tab:bus_extract}
\end{table}

\section{Counting Removed Rows}
The Data61 report provides a table of dropped trips. It shows that for LightRail, no trips were dropped for tap ons, while 0.0005\% of trips were dropped for tap offs. Looking at the data, there are 23 possible stops on the LightRail network. Additionally, there is an UNKNOWN tap off location, we assume this is where someone fails to tap off and is charged the maximum rate or where a failure in the system occurs. As such, we have 23 tap on locations and 24 tap off locations, given a total of 47. If we multiply that by the number of days, 14, we get a maximum of 658 possible rows related to LightRail. The first issue is that 0.0005\% is not a whole number of rows, it equates to $0.00329$ rows. Furthermore, when we look at the data we have a total of 658 observations. Either we misunderstand the computation or  the correct value should  be 0. 

\section{Conclusion: Evaluating Utility and Privacy Tradeoffs}
 The dataset released here does the right thing by completely breaking the links between different tap ons and tap offs by the same person.  This is critical for protecting privacy, though obviously it impacts utility by preventing any queries about trips or journeys from being answered.
It is probably inevitable that successful techniques for protecting transport customers from the leakage of  information about their trips and journeys will also limit the scientific analysis of trips and journeys.  

In general, sensitive unit-record level data about individuals cannot be securely de-identified without substantially reducing utility. Records should be broken up so that successful re-identification of one component doesn't reveal other information about the person.

The utility of a dataset depends on what queries it was intended to answer.
There will never be a zero privacy risk, so it is important that the provided utility exceeds the privacy risk. The method of evaluating the utility in Data 61's report is to consider all possible queries in the output set and calculate the error in comparison to running the query on the original data. This is an effective way of determining the impact of the perturbation on the utility of the data, but does not capture the impact of aggregation or suppression. Queries about trips and journeys cannot be answered in the released dataset.  Since both aggregation and suppression play a significant part in the privacy of the released dataset they too should be considered in the utility calculations.   

Clearly whenever a privacy-preserving methodology is applied to a dataset a significant amount of utility may be lost. The important question is whether that utility was sought or needed. Prior to planning a data release the purpose of the release, and the types of queries that it should serve, should be determined. Then the utility calculation could be determined based on the queries the dataset was intended to deliver, as opposed to those that it actually does deliver. If a dataset is being released without a specific target or set of queries to answer, further thought should be given to whether to release the dataset at all. There are finite limits on how much data can be safely released---that budget is best spent on carefully targeted releases that answer useful questions.

The authors would like to thank Transport for NSW for the opportunity to work on this project, and for agreeing to make this report public.  Openness \emph{about} data privacy is crucial for engineering  good privacy protections and earning public trust.  The minor problems and unexpected inferences we were able to identify here might help improve the privacy of future releases, by Data61 and others.  The more detail that can be published about methods for protecting privacy, the greater the likelihood that errors will be found and fixed before they are repeated.  We encourage all open data authorities to describe to the public  what they do with data, how they treat or link it, and how it is protected.

\bibliographystyle{abbrv}
\bibliography{references}

\appendix

\section{Differential Privacy Background}\label{app:dp}

Recall the definitions of differential privacy and its weaker variant.

\begin{defn}[\cite{dwork2006calibrating,dwork2006our}]
Two datasets $D, D'$ are said to be \term{neighbours} if they differ on exactly one row. 
A randomised mechanism $M$ mapping some input dataset $D$ to a vector of $d$ real numbers is said to be \term{$\epsilon$-differential private} for
privacy budget $\epsilon>0$, if for any neighbouring datasets $D, D'$ and any $T\subseteq\reals^d$, it holds that $\Pr{M(D)\in T}\leq \exp(\epsilon) \Pr{M(D')\in T}$. If for additional privacy parameter $\delta\in(0,1)$, a mechanism satisfies that for any neighbouring datasets
$D,D'$ and $T\subseteq\reals^d$, $\Pr{M(D)\in T}\leq\exp(\epsilon)\Pr{M(D')\in T}+\delta$ then $M$ is said to preserve the weaker form of
\term{$(\epsilon,\delta)$-differential privacy}.
\end{defn}

One of the first, and still most common, approaches to privatising a non-private function is the Laplace mechanism. First, we need a way to measure the sensitivity of a non-private function to input perturbation.

\begin{defn}
The \term{global sensitivity} of a (non-private) function $f$ mapping dataset $D$ to $\reals^d$ is a bound $\Delta(f) \geq \max_{D,D'} \|f(D) - f(D')\|_1$ where $\|x\|_1=\sum_{i=1}^d |x_i|$ and the $D,D'$ are taken over neighbouring pairs of data
\end{defn}

The more sensitive a target, non-private function, the more noise the Laplace mechanism adds so as to smooth out this sensitivity. A mechanism whose output is probably insensitive can preserve differential privacy.

\begin{lem}[\cite{dwork2006calibrating}]
For any (non-private) function $f$ mapping dataset $D$ to $\reals^d$ with known global sensitivity $\Delta(f)$, parameter $\epsilon>0$, the \term{Laplace mechanism}\footnote{The Laplace distribution $Lap(\v{a},b)$ on $\reals^d$ with mean $\v{a}\in\reals^d$ and scale $b>0$ parameters, has probability density function $\exp(-\|\v{a}-\v{x}\|_1/b)/(2b)$.} $M(D)=f(D)+Lap(0,\Delta(f)/\epsilon)$ is $\epsilon$-differential private.
\end{lem}

\section{Second Data Set Algorithm}
\label{app:lower}

Data61's report addendum describes The ``\alg'' (found as Algorithm 1 in Section 2, Page 4). The algorithm operates on two objects of interest:
\begin{itemize}
    \item \term{A dataset} $D$ of rows of the form $\langle d_1,\ldots,d_k\rangle$ over $k$ columns each representing common attributes such as a time, location or mode of transport. Each row represents a recorded individual trip event such as a tap on/off at a particular location and time.
    \item \term{A set of attribute combinations} we'll label $Q$, each member of the form $\langle q_1,\ldots, q_k\rangle$ with $q_i$ being a value taken from the domain of $D$ column $i$. For example, one tuple in $Q$ might represent a type of event: tap on a Manly ferry at Manly Wharf, at 06:45 on a Monday.
\end{itemize}
The goal of the \alg is to release a sanitised version of the \term{contingency table} that results from counting up, for each combination $q\in Q$ the number of matching rows in $D$. In other words, from input $D$ \alg outputs a new table with a row per $q\in Q$ and a single column corresponding to the (approximate) corresponding count. \alg achieves this goal by processing releasing an approximate count per $q\in Q$ as follows
\begin{enumerate}
\item Compute exact count $c_q(D)$ of records in $D$ matching given $q$;
\item If $c_q(D) = 0$: Release $0$.
\item Else:
    \begin{enumerate}
    \item Compute $d_q(D) = c_q(D) + Lap(0, p)$.
    \item Release $d_q(D)$ if $d_q(D)>t$ a given threshold, or $0$ otherwise.
    \end{enumerate}
    \item (Post release, the response is rounded to an integer value.)
\end{enumerate}
Threshold $t=18$ is chosen according to some reasoning about group privacy not fully explained in the addendum. $Lap(a,b)$ refers to a Laplace distribution with mean parameter $a$ and scale parameter $b$.

\begin{figure}
\begin{minipage}[t]{0.33\textwidth}
\centering
\includegraphics[width=1\textwidth]{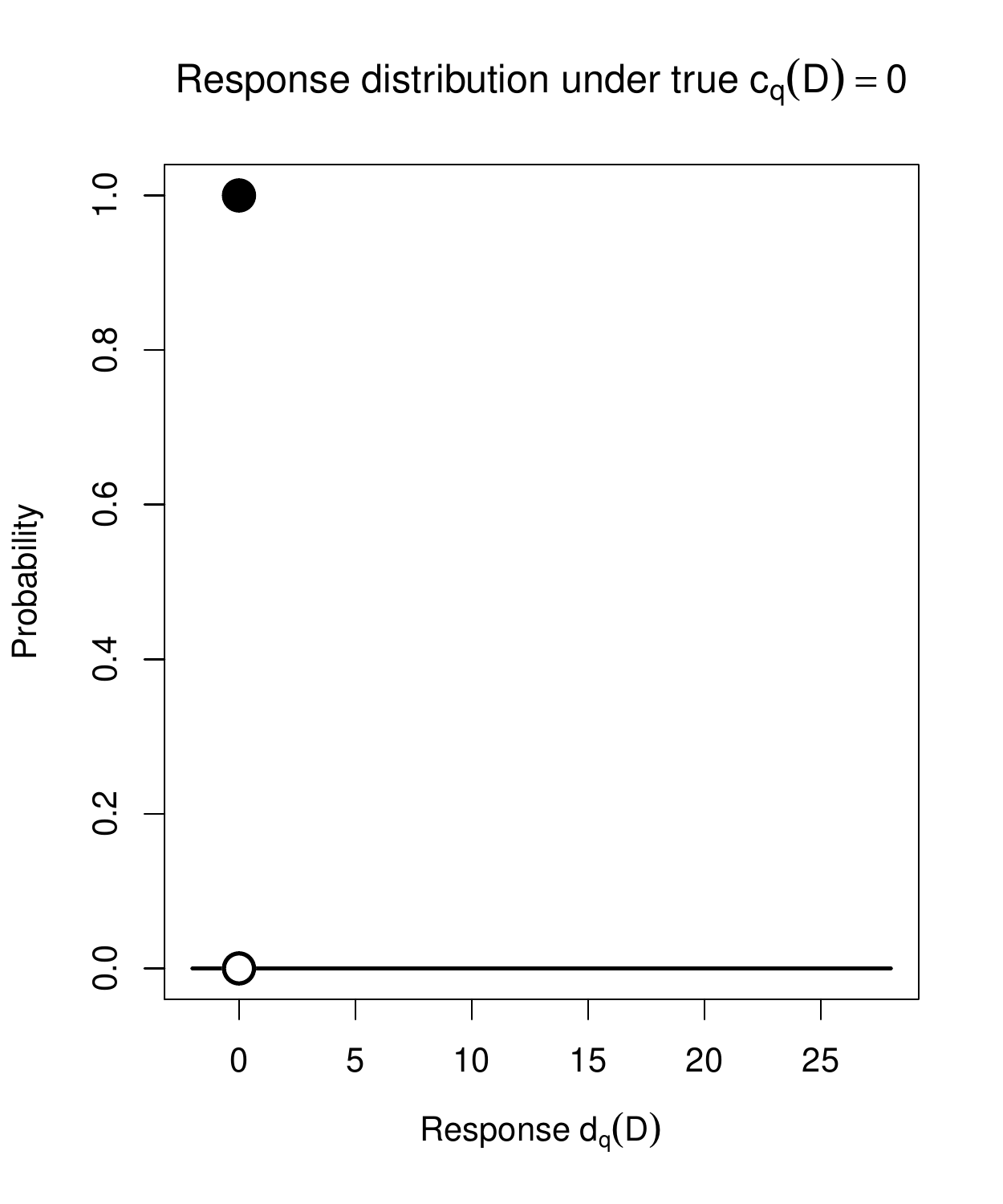} \\
(a)
\end{minipage}\hfill
\begin{minipage}[t]{0.33\textwidth}
\centering
\includegraphics[width=1\textwidth]{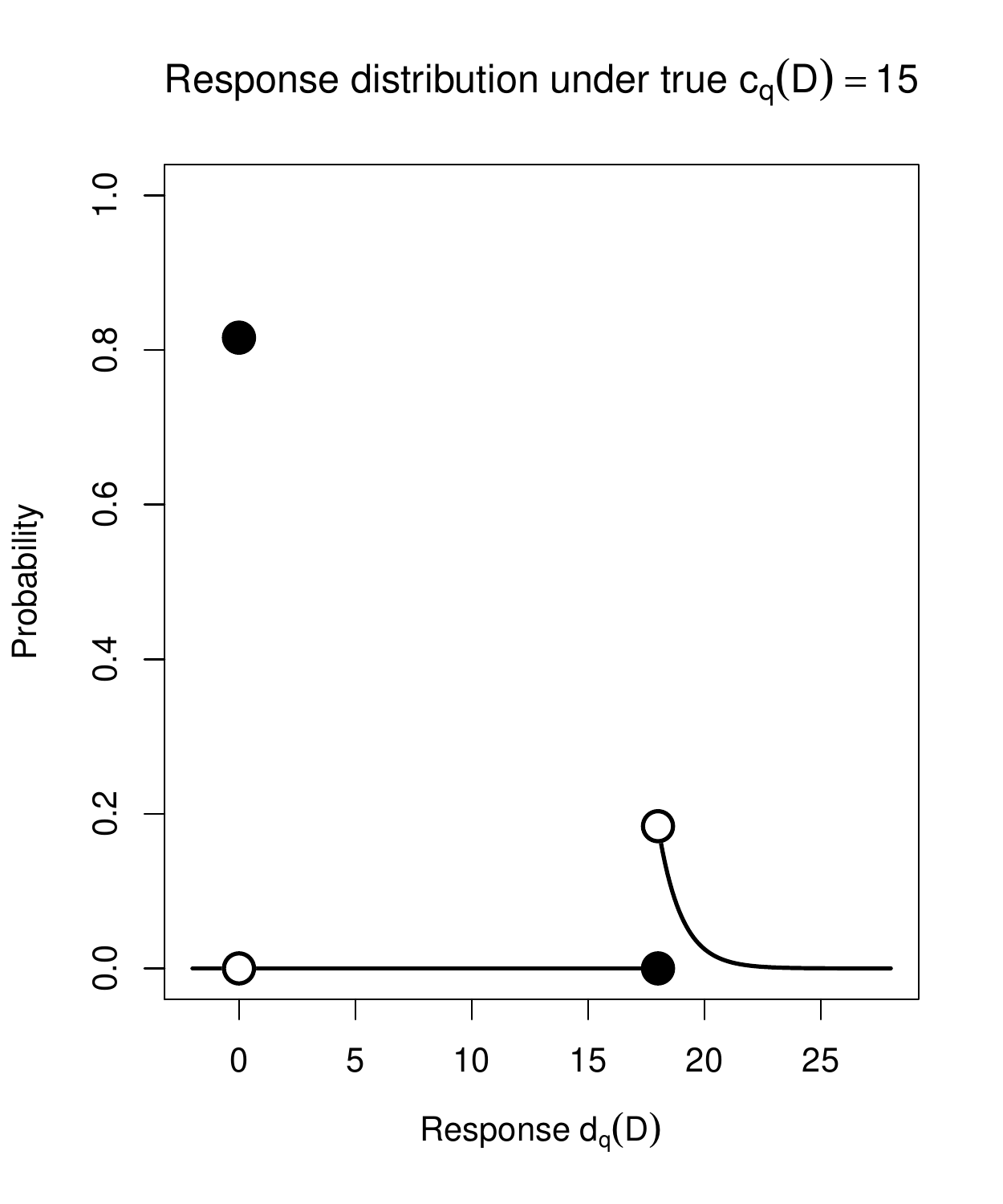} \\
(b)
\end{minipage}\hfill
\begin{minipage}[t]{0.33\textwidth}
\centering
\includegraphics[width=1\textwidth]{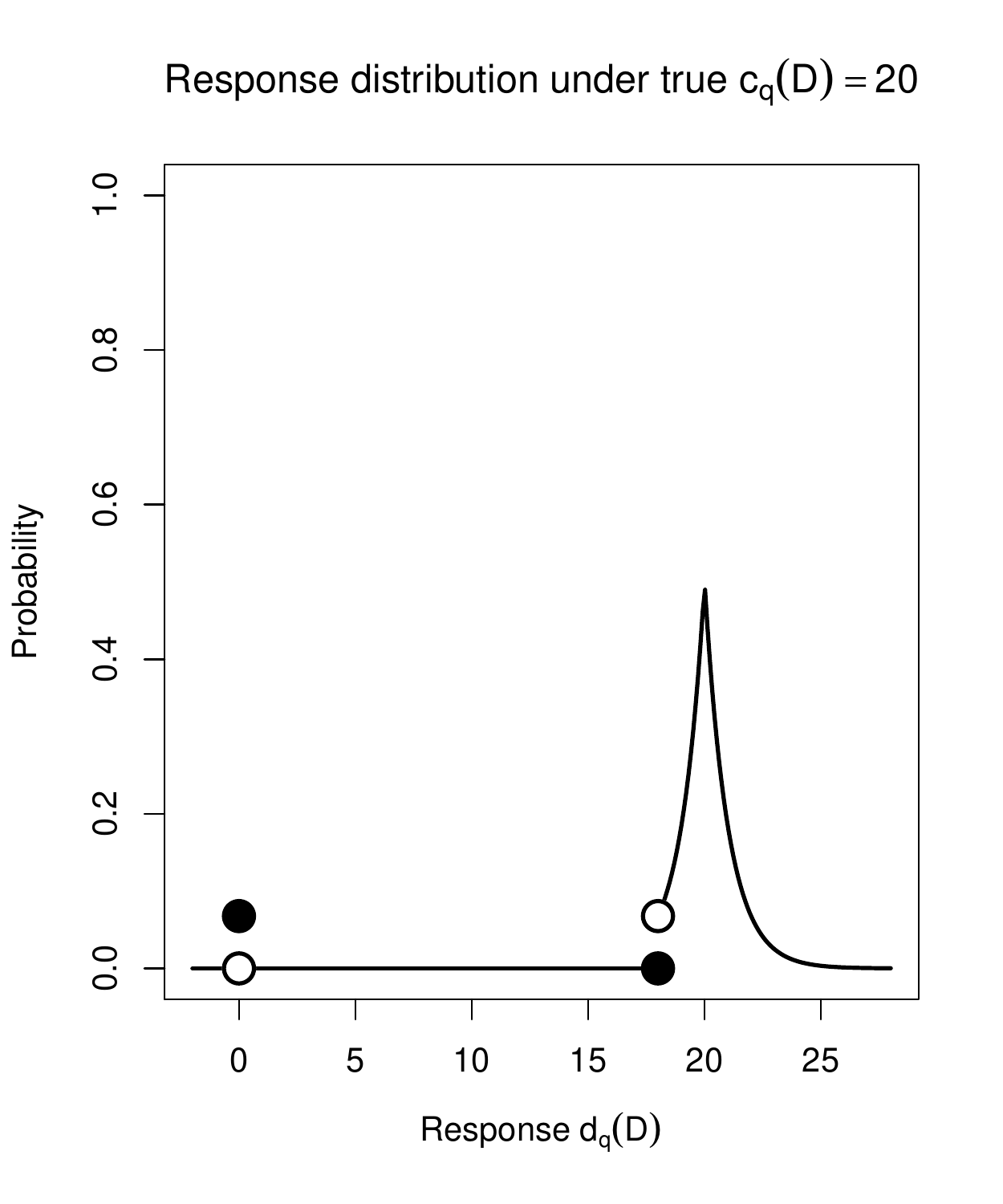} \\
(c)
\end{minipage}
\caption{The response distribution for \alg (pre-rounding) under $p=1$ and true non-private count (a) $c_q(D)=0$; (b) $c_q(D)=17$; (c) $c_q(D)=20$. Note that at discontinuities (due to suppression), the distribution takes on the value represented by the full dot; the empty dot represents the limit above/below and can be safely ignored.\label{fig:response}}
\end{figure}

Figure~\ref{fig:response} illustrates possible response distributions, corresponding to $c_q(D)=0$, $c_q(D)\in(0,t]$ and $c_q(D)>t$. These are shown pre-rounding since roudning does not affect differential privacy by a well-known post-processing lemma.

Data61's report addendum doesn't go into any detail about the type of differential privacy preserved or the level of privacy achieved (though an earlier report on a previous version of the dataset does have more detail). In the absence of a qualification, a claim of differential privacy would usually be taken to mean the stronger $\epsilon$-differential privacy.  ($\epsilon>0$ smaller being more private). 

\section{Proof of Theorem~\ref{thm:not-edp}}\label{app:not-edp}

The result follows immediately from the fact that on some datasets $D$, the \alg cannot output some values, that can be output on a neighbouring $D'$. In full detail, we need only show that there exist neighbouring datasets for which the differential privacy bound cannot hold. (Note that we are analysing the differential privacy of the algorithm pre-rounding, since rounding does not affect the level of privacy achieved.) Consider any $q\in Q$ and dataset $D$ of arbitrary size $n\in\naturals$ containing one record matching $q$, and a neighbouring dataset $D'$ with the one matching record changed to non-matching. Then for any $\epsilon>0$, focusing on the marginal response distribution $M_q(\cdot)$ on the (arbitrarily) chosen $q$,
\begin{eqnarray*}
\Pr{M_q(D)=t+1} &=& \frac{1}{2p} \exp\left(-\frac{|t+1-c_q(D)|}{p}\right) \\
 &=& \frac{1}{2p} \exp\left(-\frac{t}{p}\right) \\
 &>& \exp(\epsilon) \cdot 0 \\
 &=& \exp(\epsilon) \Pr{M_q(D')=t+1}\enspace,
\end{eqnarray*}
where the first equality follows by substitution of the PDF of the zero-mean Laplace with scale $p$; the second
follows by noting $c_q(D)=1$ by design; the last equality by the fact that $c_q(D')=0$ and so $M_q(D')$ can never output $t+1$. Note that this holds for any $t\geq 0$. The discrepancy arises from the non-perturbation of zero counts.

For general outputs of counts over a set $Q$, the same result occurs. For the joint response distribution
$\Pr{M(D)=\v{t}}=\prod_{q\in Q} \Pr{M_q(D)=t_q}$. Our construction demonstrates that for any $Q$ we can select data sets $D,D'$ such that one term in the product becomes zero, setting the whole product zero---on $D'$---while on $D$ the product will be a positive value.

The second part of the theorem follows from the same example: to achieve level $\epsilon$ on the chosen pair $D,D'$, we need $\delta$ at least the left-hand side since then
$$
\Pr{M_q(D)=t+1} = \frac{1}{2p} \exp\left(-\frac{t}{p}\right) \leq \frac{1}{2p} \exp\left(-\frac{t}{p}\right) = \exp(\epsilon) \Pr{M_q(D')=t+1} + \delta\enspace.
$$

The estimates for $\delta$ in Section~\ref{sec:DPImpossible} are simply calculations of the likelihood that the Laplace distribution produces a perturbation large enough to take the small values above the threshold of 18, assuming the parameter $p$ estimated below.

\section{Distribution of a Difference of Laplace Random Variables} \label{app:LaplaceDifferences}
This section calculates the distribution derived from the difference of two independent random variables that are distributed as 
$\textit{Lap}(0,b)$.  (What we call $b$ here is $p$ in the observed dataset.)

Let $L(b) = \frac{1}{2b} \exp({-|x|/b})$ denote the PDF of the $\textit{Lap}(0,b)$ distribution.  

For the case $u \geq 0$, the PDF of the difference between the two random variables is given by the convolution
\begin{eqnarray*}
&& \int_{-\infty}^{\infty} L(x) L (u-x) dx \\
 &=& \frac{1}{4b^2} \int_{-\infty}^{\infty} \exp({-|x/b|}). \exp({-|u-x|/b}) dx \\
&=&  \frac{1}{4b^2}\left( \int_{-\infty}^{0} \exp(x/b). \exp((x-u)/b)dx +  
 \int_{0}^{u} \exp(-x/b). \exp((x-u)/b)dx \right. \\
 && \left.+ \int_{u}^{\infty} \exp(-x/b). \exp((u-x)/b)dx  \right) \\
&=& \frac{1}{4b^2}\left( \frac{b}{2}\exp({-u/b})[\exp({2x/b})]_{x=-\infty}^{0}  + u \exp({-u/b}) - \frac{b}{2}\exp({u/b})[\exp({-2x/b})]_{x=u}^{x=\infty}   \right) \\
&=& \frac{1}{4b^2}\left( \frac{b}{2}\exp({-u/b}) - u \exp({-u/b}) + \frac{b}{2} \exp({-u/b})   \right) \\
&=& \frac{1}{4b^2}\left((u+b)\exp({-u/b})   \right) \enspace.
\end{eqnarray*}

The $u < 0$ case must be symmetric.

\begin{figure}[htbp]
\centering
\includegraphics[width=\textwidth]{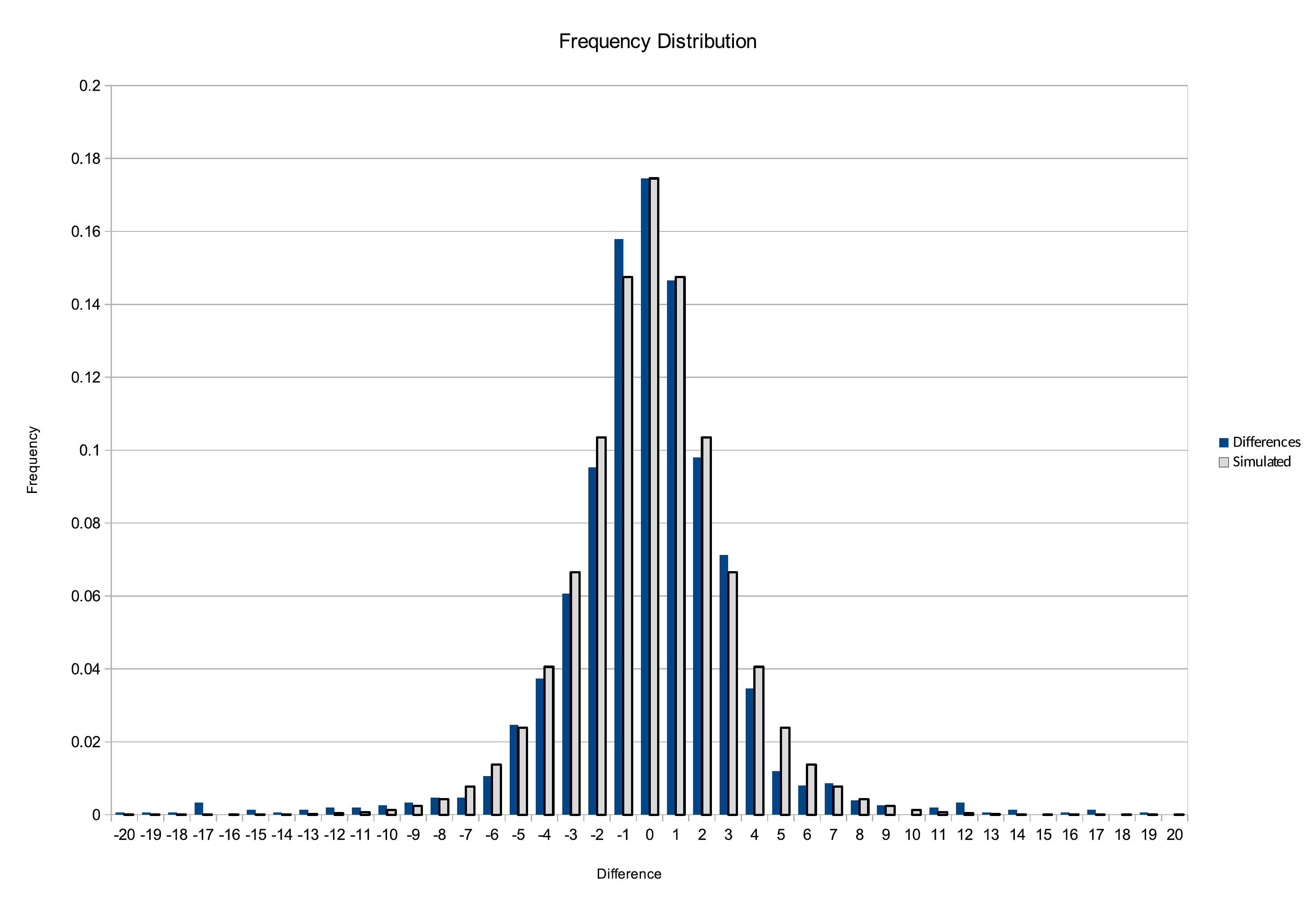}
\caption{Frequency Distribution of Manly Ferry Count Differences---Comparison of observed values against simulation with $p \approx 1.4$.}
\label{fig:freqcompare}
\end{figure}

Figure~\ref{fig:freqcompare} shows the simulation of this distribution with $p \approx 1.4$, plotted beside the observed differences from Section~\ref{sec:Recovering}.  The match is fairly close but not perfect, partly because the suppression of numbers less than 18 produces some large values in the observed data.  Nevertheless this gives us a reasonable estimate of the likelihood that these perturbations would produce large values.  

\section{Quantifying Confidence for Secret Ferries Late At Night}  \label{sec:confidenceEst}

This section leverages two indirect, privatised observations of the same trips, so as to calculate probabilities for the raw trip counts. The methodology is indicative of how differences can lead to uncovering raw passenger numbers. 

From one set of observations we have
\begin{itemize}
    \item A privatised count of 91 passengers at MFV, observing random variable $X=x+L_1$ where $x$ is the raw count and $L_1\sim Lap(0,p)$;
    \item A privatised count of 41 passengers at CQ, observing random variable $Y=y+L_2$ where $y$ is a the raw count and $L_2\sim Lap(0,p)$;
    \item A privatised count of 0 passengers at CR, based on $z\leq 18$ the raw count; and
    \item A privatised count of 150 at MFV, CQ, CR summed, observing random variable $S=x+y+z+L_3$ where $L_3\sim Lap(0,p)$.
\end{itemize}
Here the $L_i$ are independent, identically distributed. (Note that we have omitted the effect of rounding, which does not materially change the nature of the results, while it does complicate their exposition.) We have assumed that the MB ferry arrived on time and was not counted in the fourth release (this information would be easy to collect, and factor into the analysis if it arrived early). We have also assumed the same level of privacy $\epsilon,\delta$ and sensitivity for the final release as compared to the first three---based on the first report this is a reasonable assumption as it appears that only the threshold is adapted to changes in aggregation/query (the analysis should be extensible to if a different $p$ were used, although it complicates the derivation).

Now observe that the difference between the fourth release and the first three, produces a familiar quantity: a sum of independent, identically distributed random variables (all Laplacian):
\begin{eqnarray*}
S - X - Y &=& (x + y + z + L_3) - (x + L_1) - (y + L_2) \\
&=&z + L_1 + L_2 + L_3 \\
&=& \frac{1}{3}\sum_{i=1}^3 (z + 3 L_i) \\
&=& \frac{1}{3} \sum_{i=1}^3 Z_i\enspace,
\end{eqnarray*}
where the first equality follows from substituting the definitions of the random variables, the second equality follows from cancelling like terms and noting that the $L_i$ variables are symmetrically distributed about zero, the third equality shares $z$ across the three equal terms of the sum, and the final equality follows from defining independent and identically distributed random variables $Z_i\sim Lap(z, 3p)$. We denote this final average $\overline{Z}$.

We note some interesting properties of $\overline{Z}$:
\begin{itemize}
\item $\overline{Z}$ is an unbiased estimator of $z$, that is the expectation of $\overline{Z}$ is $z$. Concretely, if we were to observe $\overline{Z}$ many times we could average those observations and reconstruct $z$. In our situation we have only one observation of it, so the goodness of this ``guess" depends on the variance;
\item $\overline{Z}$ has variance $6 p^2$. By independence of the $L_i$, the variance of their sum is the sum of their variances which are all $2 p^2$ (well-known for a Laplace distribution of scale $p$). Adding constant $z$ does not change this variance;
\item $\overline{Z}$ is symmetrically distributed about its expectation. This fact isn't interesting by itself, but is used below in deriving the confidence interval. It follows from the fact that it is the average of symmetric random variables (to see this, consider that each can be replaced by its negative without affecting the distribution of the average).
\end{itemize}

Note that assuming $p$ is chosen rather small so as not to destroy utility, $p^2$ may be rather small implying just the single observation of $\overline{Z}$ is a reasonable estimate of $z$.

\subsection{A Confidence Interval}

We can also apply the standard argument for estimating confidence intervals, to this case of $\overline{Z}$ estimating $z$, included here for completeness. We deviate slightly from the typical approach, by using bounds on the distribution of the sample. The point of this exercise is to quantify uncertainty of the estimate $\overline{Z}$ of $z$.

We seek a symmetric interval about $\overline{Z}$, \eg $[\overline{Z}-a, \overline{Z}+a]$, that is likely to capture $z$, say with probability at least $1-\alpha$ for some small $\alpha$ close to zero. For example, $\alpha$ chosen as 5\% would correspond to a 95\% confidence interval. The interpretation of such a confidence interval, is that out of 100 observations of the (random) interval we expect the interval to successfully cover $z$ 95 times of the 100. This is the standard interpretation of (frequentist statistical) confidence intervals, commonly used to quantify uncertainty of estimates.

\begin{eqnarray*}
1 - \alpha &\leq& \Pr{\overline{Z}-a \leq z \leq \overline{Z}+a} \\
&=& \Pr{- a \leq \overline{Z} - z\leq a} \\
&=& 1 - 2\Pr{\overline{Z} - z \geq a}\enspace,
\end{eqnarray*}
where the inequality restates the definition of confidence interval---the interval we are seeking; the first equality follows from manipulating the inequalities/restating the condition sought; the second equality follows from the fact that the distribution of $\overline{Z}$ is symmetric about $z$.
Rearranging further, we are seeking $a$ such that
\begin{eqnarray}
\Pr{\overline{Z} \geq a + z} &\leq& \alpha/2\enspace.\label{eq:conf-int-1}
\end{eqnarray}
We have to now invert the distribution function on the left-hand side of this inequality, in order to solve for $a$. Inverting the CDF for example would yield the quantile function. In this case we don't have on hand the distribution of $\overline{Z}$ (although we could compute it, \eg using characteristic functions). Instead we will bound the probability in terms of a probability that is known and easily invertible. Using a bound does not invalidate the correctness of the confidence interval derived, but rather will produce a wider interval than necessary; as we shall see, the interval is sufficiently tight for our purposes. A natural way to bound tails of i.i.d. sample sums/averages is to use concentration inequalities such as Hoeffding's inequality. We don't bother pursuing such approaches since we have a sample of such a small number of random variables (three): a simpler/looser route works well.

Note that the following events' relationship
\begin{eqnarray*}
\bigcap_{i=1}^3 \{Z_i \geq z + a\} &\Rightarrow& \{\overline{Z} \geq z + a\}\enspace,
\end{eqnarray*}
implies a bound on our probability of interest in~\eqref{eq:conf-int-1}
\begin{eqnarray*}
\alpha/2 &\geq& \Pr{\overline{Z} \geq z + a} \\
&\geq& \Pr{\forall i,\ Z_i\geq z + a} \\
&=& \prod_{i=1}^3 \Pr{Z_i\geq z + a} \\
&=& \prod_{i=1}^3 \frac{1}{2}\exp\left(-\frac{a}{3p}\right)\\
&=& \frac{1}{8}\exp\left(-\frac{a}{p}\right)\enspace,
\end{eqnarray*}
where we have substituted in the known cumulative distribution function for $Z_i\sim Lap(z, 3p)$, leveraged the independence of these variables, and simplified. Solving for $a$ yields
$$a \geq p \log\left(1/(4\alpha)\right)\enspace.$$
Summarising, we have proven the following result.

\begin{thm}\label{thm:confidence-interval}
A confidence interval for the actual count $z$ of passengers at CR, of confidence level $1-\alpha$ for any $0<\alpha<1$, is given by 
$S-X-Y \pm p \log\left(1/(4\alpha)\right)$.
\end{thm}

Given the observed $X,Y,S$, if $p=1.4$ for example as estimated above, then we estimate that
$z$ is in the range 17.02 to 18.98 with confidence 95\% (capturing only one integer 18). A 99\% confidence interval is 16.04 to 19.96 (capturing integers 17, 18, 19).

\end{document}